# Forensic Artefact Discovery and Attribution from Android Cryptocurrency Wallet Applications

**Eugene Chang[1,2], Paul Darcy[2], Kim-Kwang Raymond Choo[3] and Nhien-An Le-Khac[2]**
[1]York Regional Police, Ontario, Canada
[2]School of Computer Science, University College Dublin, Ireland
[3]The University of Texas at San Antonio, San Antonio, USA

Corresponding author: Nhien-An Le-Khac (e-mail: an.lekhac@ucd.ie).

**ABSTRACT** Cryptocurrency has been (ab)used to purchase illicit goods and services such as drugs, weapons and child pornography (also referred to as child sexual abuse materials), and thus mobile devices (where cryptocurrency wallet applications are installed) are a potential source of evidence in a criminal investigation. Not surprisingly, there has been increased focus on the security of cryptocurrency wallets, although forensic extraction and attribution of forensic artefacts from such wallets is understudied. In this paper, we examine Bitcoin and Dogecoin. The latter is increasingly popular partly due to endorsements from celebrities and being positioned as an introductory path to cryptocurrency for newcomers. Specifically, we demonstrate how one can acquire forensic artefacts from Android Bitcoin and Dogecoin cryptocurrency wallets, such as wallet IDs, transaction IDs, timestamp information, email addresses, cookies, and OAuth tokens.

## I. INTRODUCTION

Since 2012 cryptocurrency wallets have reportedly grown from 10 thousand users worldwide to over 81 million users in March 2022 according to Statista.com (blockchain wallet users worldwide)[1]. Figures from Statista.com also illustrate that the Android market share worldwide has grown from 25% in 2012 to 73% in 2021. This trend reflects rapid growth in technological advancements that law enforcement agencies must adapt to when conducting their criminal investigations. Cryptocurrencies such as *Bitcoin* and *Dogecoin* rely on blockchain protocol, privacy, and its cryptography. With the advancements and adoption of digital currencies, cryptocurrency application wallets are not usable without the value they hold in the form of private keys [1].

The *Bitcoin* protocol is the first cryptocurrency based on a decentralized system and is synonymous with being the face of cryptocurrency. Many variations have been derived from Bitcoin and one of them is Dogecoin, with an almost 'cult-like' status that has its beginnings as a 'joke' cryptocurrency coin. *Coinmarketcap.com* shows its growth in value from close to zero in February 2021 to almost .75 cents USD in May of 2021[2]. The recent surge and popularity of Dogecoin has made it a popular topic in mainstream media and despite its volatile price, it has become the current poster child for cryptocurrency at the time of the writing of this research paper and had little to no research conducted on this coin.

A perspective by Interpol on forensic challenges with Cryptocurrencies did mention in [2]. Traditional criminal investigations have historically depended on physical evidence, but with the recent advancement and increasing popularity of cryptocurrency technology, law enforcement is facing the need to be able to draw on digital evidence. Whereas electronic payment systems used to contain fiat cash, credit and debit card transactions, payments are increasingly transitioning to electronic cryptocurrency wallets. These cryptocurrency transactions are increasingly being used in crimes such as drug trafficking, murder, assault, and child pornography. With cryptocurrency technology designed to disguise and protect the

---

identity of its users, law enforcement agencies need to find ways to extract digital evidence from these highly encrypted transactions as laid out by [2].

However, law enforcement agencies have not taken a greater advantage of cryptocurrency forensic evidence that may be present in mobile devices such as Android smart phones. For example, digital forensic artefacts exist on mobile devices that contain cryptocurrency wallet applications, but these artefacts are not commonly being extracted for use in criminal investigations.

In most existing criminal investigations, we generally require a broad range of tools and techniques to collect large volumes of cryptocurrency and related evidence. In other words, there is no simple forensic sought and efficient approach for the acquisition and analysis of evidence from cryptocurrency wallet applications. This is a significant challenge to regional police agencies with limited budgets; thus, reinforcing the importance of designing simpler efficient methods to collect digital evidence. In the context of Android devices, for example, Cookies and OAuth tokens can enhance user experience by storing credentials and sessions, so that a user does not have to constantly re-login to installed applications. We posit the potential of using Cookies and OAuth tokens to facilitate forensic investigations. However, the hardened cyber security measures in newer operating systems and Android devices may complicate the acquisition of forensic artefacts. From our literature review, there has been no research focused on web-based cookies and OAuth tokens in the context of cryptocurrency wallet forensic analysis.

This research aims to propose a new method to identify human readable digital forensic artefacts associated with *Bitcoin* and *Dogecoin* transactions on Android wallet applications, *Coinbase v.1.22.3, Coinomi v.9.26.3* and *Atomic v.0.75.1*, in order to facilitate attribution during a criminal investigation. A summary of our work in this paper is as follows:

- We present a process to acquire and analyze the forensic extractions of artefacts from *Bitcoin* and *Dogecoin* cryptocurrency exchange and wallet applications installed on Android devices.
- Using the process, we demonstrate how one can identify important artefacts (e.g., Hash IDs, addresses, cookies, and OAuth password artefacts) and conduct examinations to support attribution such as exchanges that facilitated these transactions.

The rest of this paper is organized as follows: Section II provides background concepts related to Bitcoin and Dogecoin, cryptocurrency wallet and transaction models. This section also mentioned the ability of using Cookies and OAuth tokens as forensic sources. Section III discusses the related cryptocurrency forensics approaches. In Section IV, we introduce our proposed cryptocurrency forensic approach prior to demonstrating how the approach can be used to identify and extract relevant digital forensic artefacts of cryptocurrency transactions on Android devices in Section V. Finally, Section VI concludes this paper. Fig. 1 is the flowchart that describes the relationship between the various sections.

## II. BACKGROUND CONCEPTS

### A. BITCOINS and DOGECOIN
Bitcoin is the original cryptocurrency developed by the mysterious character Satoshi Nakamoto in 2009 and as of July 2021 holds the highest market-cap of almost 700 billion dollars USD relative to all cryptocurrencies according to coinmarket.com. Currently, Bitcoin can be purchased by most cryptocurrency exchanges and is known as the most common cryptocurrency in the market.

However, not much is known about Dogecoin, apart from the online popularity generated by people such as Elon Musk who quoted, "Working with Doge devs to improve system transaction efficiency," causing the value of Dogecoin to increase by almost 40%[3]. Although there are concerns about the efficiency of POW cryptocurrencies, Dogecoin has its site on Defi (Decentralized Finance) that is another massive topic of discussion that revolves around revolutionizing the traditional financial system. This move that Dogecoin forecasts for its own technology will promote faster, cheaper, reliable and more secure transactions as mentioned in [3]. Dogecoin has been somewhat of an internet phenomenon, yet there is little to no research about this cryptocurrency.

According to [3], Dogecoin is a Proof-of-Work implemented cryptocurrency that was created in 2013 by its founders Bill Marks and Jackson Palmer. This cryptocurrency is a fork from the Litecoin cryptocurrency which according to [4] is a "new independent software project" copied from an original project and in this case, Litecoin[4]. Litecoin's unique features apart from Bitcoin is that it obtains faster transactions speeds that Bitcoin. The maximum number of Bitcoins that are allowed on the network are 21 million coins, whereas Litecoin is 84 million which likens itself to silver, if Bitcoin is gold.

### 1) DOGECOIN DIFFERENCES AND SIMILARITIES TO BITCOIN

---

There are no academic resources found for the structure of Dogecoin, but since the nature of cryptocurrency is decentralized and there is a large community that supports the operations of Dogecoin, open-source documentations are valuable in understanding Dogecoin.

As seen in the guru99.com[5], the differences are quite different. Dogecoin uses the Scrypt algorithm, which is a hashing algorithm that, "takes large amount of memory and cpu cost" as mentioned in [5]. The Scrypt algorithm also is secure against brute-force attacks than other alternative hashing algorithms and uses less energy than other alternatives as mentioned in [6]. Although there are advantages to the security ecosystem of Dogecoin in comparison to Bitcoin, it is forecasted that Proof of Work coins like Dogecoin will migrate to a more environmentally friendly technology as mentioned in Dogecoin's whitepaper.

Unlike Bitcoin public addresses, Dogecoin addresses that represents the public key starts with the letter 'D' and would look similar to this: **DFMpT5Q4PCrfckQptRYtyvGmZ2GKoYANNH[6]**.

As of July 2021, *Dogecoin* is available to trade and purchase at a variety of cryptocurrency exchanges. The exchange that provides the largest trade percentage of *Dogecoin* is Binance at 43.3 %.

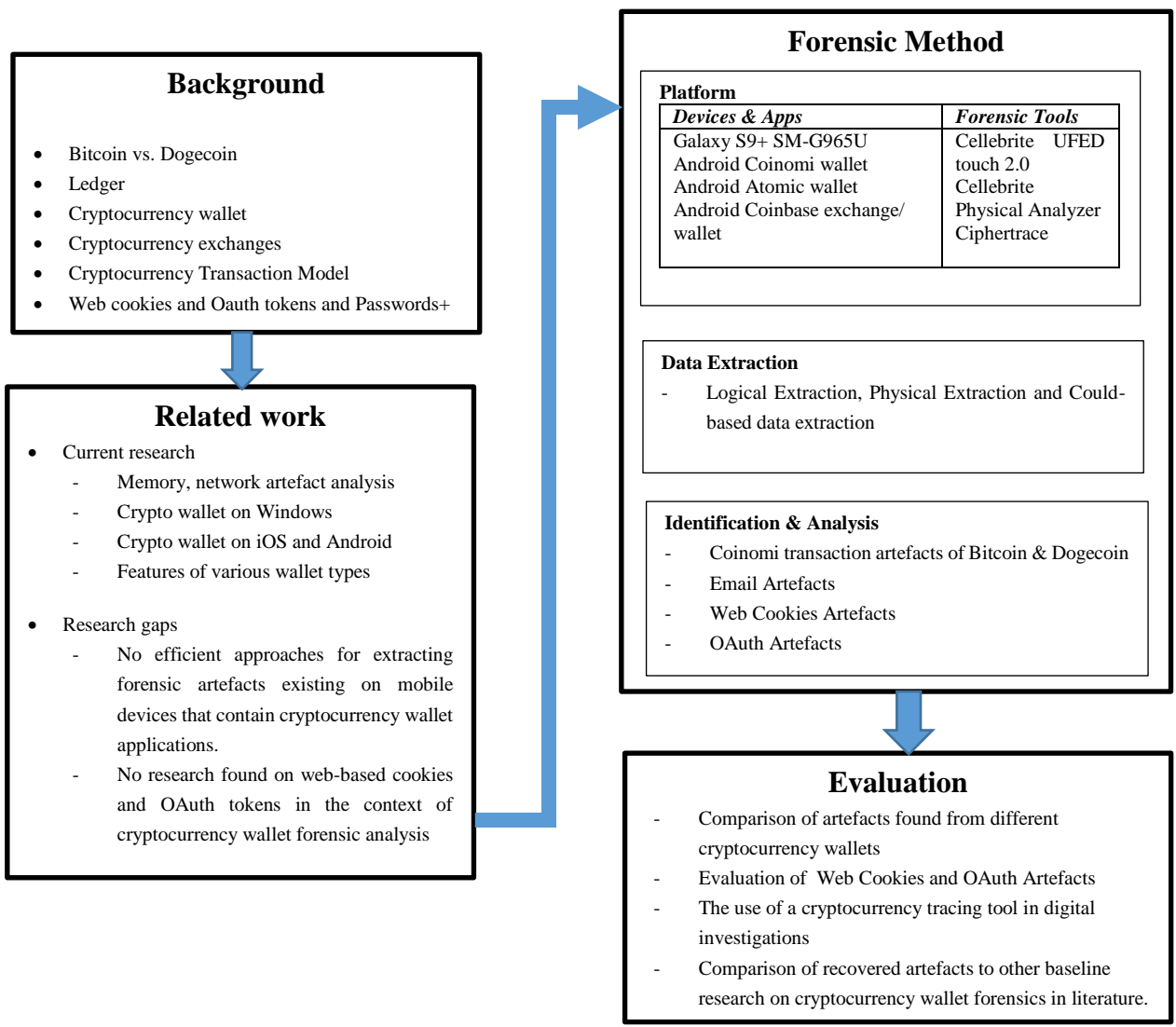

**FIGURE 1.** The paper flowchart: Relationship between main sections

## B. LEDGER

A traditional ledger is a familiar concept that has been around for quite some time. It predates the mass adoption of digital technology when writing utensils and paper were a common form of documentation. For example, in the 1700's during the colonial period in the United States of America, a ledger was used to record household goods held on credit.

A modern understanding of a ledger in the context of criminal investigations is the use of a 'debt list' or a ledger used by drug traffickers that document their drug trafficking operations as shown by [7]. A drug trafficker could keep a list of drug deliveries while noting the date, weight and value.

Much like the previous examples of ledgers, the concept has been modernized and adopted within blockchain technology. A ledger, in the context of blockchain holds the same principles where each blockchain holds a chain of blocks that contain information into groups that cannot be altered. Inside each block contains data and a hash key for the current block and one from the previous block forming a ledger.

What makes the blockchain an alluring technology is that it embraces characteristics of decentralization and security that prevents the alteration of data contained in the blocks. In other words, blockchain technology is a type of a distributed ledger.

## C. CRYPTOCURRENCY WALLETS

In order to interact with cryptocurrency, software also known as a cryptocurrency wallet, by accessing the internet is required to initiate transactions on the blockchain distributed network. Suratkar, S. et al. [8] describes that cryptocurrency coins, unlike traditional fiat currency, is not a digitally tangible asset stored on a wallet.

A wallet allows access to the blockchain and the ownership is verified and secured by a cryptographic key stored in the wallet. The classification of these types of wallets are identified into three categories: paper, hardware and software.

They also described a mobile wallet as an application on a mobile device, commonly available on Android and iOS operating systems, that stores cryptocurrency data [8]. A mobile software wallet can also be called a 'hot wallet' that has a permanent connection to the internet, which allows it to interface with the blockchain.

For a user to back up a wallet in the event the device is lost, stolen or seized, a pneumonic seed phrase would be assigned to the wallet address that can access the wallet without a passphrase as mentioned in [9] that would look something like: WITCH COLLAPSE PRACTICE SHAME OPEN FEED DESPAIR LEAST ICE AGAIN ROAD CREEK

Some applications would offer the user the number of words to choose for a phrase. Since it is difficult for most people to remember these phrases, criminals would need to decide on how to store these phrases. Some of the options available would be to write them down on a piece of paper and store them in a secure location. This would be practical, but a consideration needs to be made about how safe it is since something like this can be easily misplaced or damaged to the point of it being unreadable. Another option would be to have the phrase engraved into metal and stored and a secure location such as another address or safety deposit box. A convenient way of backing up a seed phrase would be to copy and paste it in a document on the host device. These are some factors that investigators will need to consider for cryptocurrency artefacts.

There are two keys stored inside a mobile software wallet: private key and public address [8]. For a transaction to be completed, the private key must sign the transaction and be associated to the public key which is shared openly.

Although all cryptocurrency addresses are a hexadecimal string, each cryptocurrency coin has its own unique distributed ledger and their formats are unique to their technology. For example, Yessi, B.P., [10] explains that public Bitcoin addresses are created to be used only once for each transaction and that they are generally made up of 26-35 hexadecimal characters that can also be represented as a QR code. Below is an example of a paper wallet that shows both public address and private key for a Bitcoin, displaying QR codes that can be captured by cameras within a wallet application.

This would prove to be a challenge for Investigators if a variety of address artefacts are to be attributed to one individual of interest in criminal activity. Wallet addresses are valuable forensic cryptocurrency artefacts that an Investigator can look for while conducting an investigation.

*D. CRYPTOCURRENCY EXCHANGES*

Franco, P. [4] indicates that cryptocurrency exchanges are entities that convert cryptocurrency to fiat currency. This allows a person to be able to trade fiat to cryptocurrency and trade between different digital assets. Three types of exchange services are identified in [11] and are identified as:

1) Order-book exchanges: Platform that uses a trading engine to match buy and sell orders from users.
2) Brokerage service: Service that lets users conveniently acquire and/or sell cryptocurrencies at a given price.
3) Trading platform: Platform that provides single interface for connecting to several other exchanges and/or offers leveraged trading and cryptocurrency derivatives.

Although dedicated cryptocurrency wallet applications were installed on the test device for this paper, the Coinbase exchange Android application (a large exchange that offers the services mentioned above) was used and can also send and receive cryptocurrency transactions. This can be seen in the description of the results. Certain dedicated wallet applications, such as Coinomi, can also connect its users to cryptocurrency exchanges to purchase cryptocurrency with fiat money.

*E. UTXO (Unspent Transaction output)*

Franco, P. [4] explained a UTXO as a ledger on the blockchain that records unspent transactions outputs. A common mistake an Investigator may make when examining cryptocurrency forensics on a mobile device is the inability to delineate between the wallet addresses, hash (transaction IDs) and Unspent Transaction Outputs (UTXO). There are a lot of parallels between traditional fiat transactions and cryptocurrency transactions and for example, UTXO's are the change that you would receive back into your wallet from making a transaction.

*F. CRYPTOCURRENCY TRANSACTION MODEL*

One can understand this by making a parallel between fiat transactions and cryptocurrency transactions (Table I):

TABLE I
TRANSACTION EXAMPLES

| Fiat Example | Cryptocurrency Example |
|---|---|
| Leather pocket wallet | Android software (hot) wallet |
| US, CDN, Euro bills in your wallet | Different cryptocurrency coins such as Bitcoin and Dogecoin. |
| Printed receipt that shows the transaction of a paid good. | Hash or Transaction ID of a cryptocurrency transaction |
| Change from a transaction and returned back into leather pocket wallet. | Unspent Transaction Output (UTXO) |

As a visual example (Fig. 2), JIM has 5 Bitcoins and holds a Coinomi Android Bitcoin Wallet. ANNA has 1 Bitcoin and holds an Atomic Android Bitcoin Wallet. When JIM sends ANNA 1 BTC, JIM's would open his wallet and input/scan ANNA's wallet address and chooses to send her 1 BTC. At this time, JIM's output address would be signed with his private key along with ANNA's address 1 and generate a transaction hash. This would include the input and output addresses along with a date/time stamp. The transaction would be recorded on the Bitcoin ledger that JIM's address 1 sent ANNA's address 1, 1 BTC on a particular date.

In addition, it is important to note that the output of 2 BTC from JIM's address would equal the input of 2 BTC, which is comprised on ANNA's receipt of 1 BTC and the input of the UTXO back into JIM's wallet.

Wallet types, input and output addresses, UTXOs and transaction IDs would be common artefacts an Investigator would come across during a forensic examination.

Now if JIM sent 1 BTC to ANNA and 1 BTC to JON's Coinbase wallet, both outputs would be signed by JIM's private key and create unique hashes for both transactions.

Again, the output would include the input of BTC transacted and in this case when JIM sent JON 1 BTC, JIM's address 1 sent out 3.5 BTC where 1 BTC was sent to JON's address and 2.5 BTC was unspent (UTXO), sent as an input back to JIM's wallet (Fig. 3).

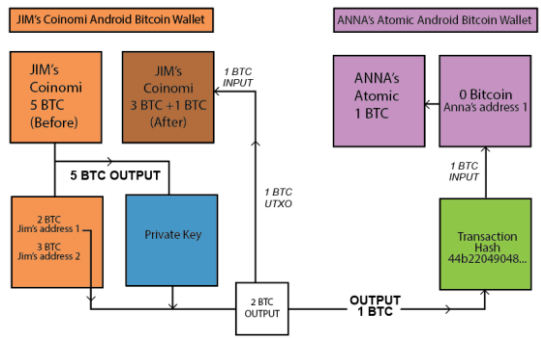

**FIGURE 2.** UTXO from single transaction

In the transaction between JIM and ANNA, JIM's address 2 sent out 1 BTC to ANNA's address where 0.5 BTC was unspent (UTXO), sent back as an input back to JIM's wallet.

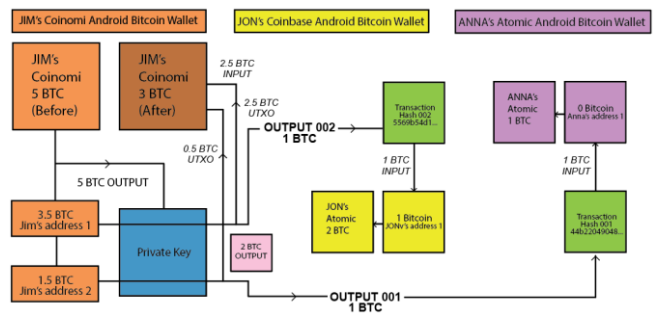

**FIGURE 3.** UTXO from two transactions

### G. WEB COOKIES AND OAUTH TOKENS AND PASSWORDS+

Software and computing have a historical relationship with Software as a Product (SAAP) also known as a 'classical perpetual license' as mentioned in [12], but also mentioned in [13], there has been a shift towards Software as a Service (SAAS) as the SAAP model is appealing because of its continued access to modernization and updates.

Understanding this narrative, software licenses require verification for use and though cryptocurrency prides itself with decentralization and privacy, Android and other mobile OS platforms battle to create a better user experience while maintaining a high level of security.

A common form of authentication on the Android OS is the use of web Cookies and OAuth token authentication. These are artefacts that an investigator may come across while examining an Android device, yet there has not been a lot of forensic analysis seen on these items.

Hypertext Transfer Protocol (HTTP) allows the communication between a web server and a web browser (client/user agent). When clicking on a hyperlink on a web browser, a user is directed to a web server for that request. LaCroix, K et al. [14] explained that a cookie is information passed back and forth between a web browser and web server providing the web server with a 'memory' of where it 'left off' so that the process of logging in is doesn't need to be done every time a user visits the website/application.

Lee, W.B. et al. [15] explained that HTTP cookies, "are used to store user-related information sent by a website, and they can be read again later to maintain a link between a user's computer and the website and to remember the user's previous state on the website. In cloud services, cookies are used by service providers to maintain smooth operation for users." Cookies are vulnerable when they are being transmitted over the internet. For example, they can be eavesdropped on when sent over the internet and locally, they can be "maliciously altered" to "manipulate the web server".

In the context of cryptocurrency wallet applications installed on an Android device, a consideration can be made to understand the behavior of cookies based on its vulnerabilities.

Shehab, M et al. [16] conducted a security analysis on different OAuth functions on smart phones. Mobile applications generally require access to resources that are stored on resource servers where authentication servers grant credentials in order to access the application. Security flaws can exist if a sophisticated adversary can implement malicious code on a device that provides connecting services to applications for the purpose of collecting user data.

Yang, F. et al. [17] describes the OAuth 2.0 protocol as a "generic framework to let a resource authorize third party access the owner's resource held at a server without revealing to the third-party the owner's credentials (such as username and password)." Although this paper looks into digital artefacts found inside a mobile device, conducting an investigation into cryptocurrency for evidence of money laundering or proceeds of crime, requires additional judicial authorization to access wallet accounts.

## III. RELATED WORK

Koerhuis et al. [9] provides a modern perspective on privacy-oriented cryptocurrencies such as Monero and Verge coins. Privacy-oriented cryptocurrencies are a valuable object of investigation, since criminals require obfuscation to commit their crimes and 'cover their tracks'. This research was based on analyzing these cryptocurrencies on a Macbook personal computer running VMware software with an approach to acquire forensic artefacts through researching memory, capturing network data and analyzing the disk. The advantage of this research provides insight into the memory, disk and network artefacts from a laptop computer that would be beneficial to a criminal investigation in addition to attribution but does not provide insight into a mobile device such as an Android phone. According to Statcounter.com, mobile phone users exceeded computer users and currently have almost 15 percent more users today over computer users. Research of new mobile applications conducting cryptocurrency transactions are not analyzed in this document which would not look in the larger demographic of mobile phone users.

Van Der Horst, L. et al. [18] conducted research on Windows based cryptocurrency wallet clients. Memory analysis is detailed and precise strategy in examining cryptocurrency wallet artefacts that identified transaction/hash IDs in the Windows operating system, but the technical skills and knowledge may not be readily available to a common Investigator that may be extracting and analyzing the high intake volume 10-20 of devices per Investigator a day. Identifying the same cryptocurrency artefacts in an Android file system would contribute to the abilities of a law enforcement Investigator and may reduce the time spent on a particular device.

Li, X. et al. [19] analyzed four cryptocurrencies based on their popularity and highest market cap. Bitcoin, Ethereum, XRP and Libra were the subject of this research that discussed the features of these coins that included how their identity was managed, consensus algorithms and coin supply. The advantage of this research is modern and provides insight into popular and valued cryptocurrencies, but as of August 2021, the top three cryptocurrencies with the highest market cap do not include XRP and is replaced by Tether and Binance coin. This is indicative of the rapid change and volatility of cryptocurrency growth that would require research to be modern. This research also did not conduct analysis into attribution or the behavior of the subject cryptocurrencies for attribution.

Montanez, A., [20] conducted research on cryptocurrency wallet applications on both iOS and Android mobile devices for forensic artefacts. This is a relevant topic of research that has a similar purpose to this paper but was conducted in 2014. Three coins were chosen for examination that included Bitcoin, Litecoin and Darkcoin. A particularly interesting review to this research is that Darkcoin has been rebranded as Dash which no longer trades on the Darkweb as it did previously. A common theme appears is that the landscape of cryptocurrency is volatile where what was once popular, can disappears as fast as it was created. The tools used are also similar to the ones used in this research, however, the technology and versions of software has changed significantly in order to adapt to more modern mobile devices. One in particular is that it does not provide add-on tools such as Cloud Analyzer which is discussed in this research. A method in this research used a virtual Android device, which may have been beneficial if a network analysis was conducted in this research, but with the advancements in Cellebrite technology it would not prove to be as beneficial as it was in 2014. SQLite databases were also analyzed in Montanez [20]'s research which was not conducted in this research since the objective is to examine human readable data. This research will add to the literature by providing tracing analysis of artefacts that are found in modern Android devices using recent tools.

Suratkar, S. et al. [8] conducted research that reviewed the features of various wallet types such as desktop, online, mobile and hardware wallets. The review of the mobile wallet provided insight into its basic theories that proved beneficial in understanding the similarities between competing wallets in the Android operating system. For example, features of cryptocurrency wallets such as key management, anonymity and exchanges are some that may be more timeless than a cryptocurrency itself. The knowledge of these features can be applied to future cryptocurrency wallet examinations. In particular, the research stated that Coinomi does not accept fiat currency, but in only one year, the Coinomi wallet application has third party features that now accept fiat currency to receive various types of cryptocurrencies to its wallet in Android. This feature is discussed later in this paper.

Nabi, A.G., [21] research objective was to evaluate three different types of cryptocurrency wallets to consider the technical and theoretical aspects on an Android device. In particular, Nabi reviewed the same Android Coinbase wallet application researched in this paper. Much of the features of this application have not changed, but the research did not examine the forensic artefacts that would be found on an Android device.

As indicated previously, much of the literature for cryptocurrency and wallet forensics can be outdated quickly. For the user, cryptocurrency is a volatile investment with prices ranging from pennies to dollars in months and vice-versa that would influence an Investigator to adapt to these changes by taking a modern approach to cryptocurrency research. As a law enforcement Investigator, the goal of attribution is essential to investigating cryptocurrency related criminal investigations. Much of the research mentioned previously provides valuable technical characteristics and reviews of cryptocurrency and their wallets, but do not provide attribution past identifying cryptocurrency artefacts.

Baksh, M. [22] discusses the difficulty of finding attribution in investigating cybercrime. To obtain indictments, an Investigator would require a "high level of attribution" than in other areas such as the intelligence community that provides corroborated information that may direct an investigation. The environment being discussed considers the area of the dark web where criminal can obfuscate their identity online. In this paper, the environment being investigated is the realm of physical devices that would generally be seized in during an investigation. This journal is beneficial to the intent of this research as it prioritizes the necessity of finding attribution, but what it doesn't provide is a technical solution to acquire attribution.

General research into cryptocurrency forensics is broad in scope, but human readable evidence useful to criminal investigations is lacking. Current techniques and methodologies are valuable in acquiring forensic artefacts, but they may be difficult to understand in a judicial context. There is a need for more simplified methods of collecting and presenting digital evidence for law enforcement.

Therefore this research focused on a new method of acquiring prosecutable evidence of cryptocurrency found on Android devices using tools available to law enforcement. Related work on this topic provided some valuable evidence using techniques that may be too sophisticated for law enforcement. It is not only wallet and transaction IDs embedded in emails were identified, but also a deeper analysis of other artefacts such as web cookies and tokens could provide investigators with user credentials that would expose further attribution. So far, such approaches with the ability to conduct this type of analysis have not been readily available to law enforcement yet.

## IV. FORENSIC METHODOLOGY

### A. PLATFORMS

To simulate a criminal device for examination, popular cryptocurrency wallet and exchange applications were installed on an Android test device prior to extraction through the Google Play Store and the email address, which set up these accounts was kept on the device. Digital forensic tools available to law enforcement agencies were used for the purpose in preparing, identifying, and analyzing the cryptocurrency forensic artefacts found on the Android test device.

TABLE II
DEVICE AND APPLICATIONS

| Device and Applications | Version |
| --- | --- |
| Samsung Galaxy S9+ SM-G965U phone | Android 8.0 Oreo |
| Android Coinomi wallet | 9.26.3 |
| Android Atomic wallet | 0.75.1 |
| Android Coinbase exchange/wallet | 1.22.3 |

TABLE III
FORENSICS TOOLS

| Tool | Version |
| --- | --- |
| Cellebrite UFED touch 2.0 | 7.45.0.96 |
| Cellebrite Physical Analyzer | 7.45.1.43 |
| Ciphertrace | June 2021 |

A Samsung mobile device was chosen for this examination since it is one of the most popular Android devices with the Samsung having the greatest market share of all Android providers. Mobile phone models vary according to the geographic area they are to be released at. A Samsung Galaxy S9+ phone (model S9+ GSM SM-G965U) is used in our experiments. The test device running Android 8.0 Oreo with no SIM card attached was rooted. The successful rooting of the test device provides the most robust data for extraction and analysis to provide attribution (Table II).

Regarding the digital forensic tool (Table III), *Cellebrite UFED touch 2.0 v. 7.45.0.96* was chosen. Three types of acquisitions were obtained while extracting data from the test device: Physical, File System and Logical extractions. To acquire a physical extraction, the Android bootloader was unlocked, and a custom recovery image was installed. A separate attempt to acquire a physical extraction on an unrooted test device proved to be unsuccessful and without a physical extraction, there were some relevant forensic artefacts that were missing, which will be explained in the results of the experiment. A logical acquisition provides a bit-by-bit acquisition of the logical memory at current state of the device.

Three popular cryptocurrency applications were chosen for the experiments based on its popularity on the Google Play Store at the time of writing this paper: *Coinomi Wallet v.9.26.3, Atomic Wallet v.0.75.1 and Coinbase exchange v. 1.22.3*, a trading platform for cryptocurrency. A general understanding of the features for each application will be documented such as their domains, IPs and name servers to identify any network artefacts found in the extraction of the test device. In addition to the features of each application, a review of how each application was setup for use will be documented to compare any artefacts found in the application from the setup.

### B. DATA EXTRACTION

*Samsung S9+ GSM SM-G965U* test device was connected to a *Cellebrite UFED Touch 2.0 v. 7.45.0.96*. While connected to the UFED Touch, the test device model is supported and provides Advanced Logical, File System and Physical data extractions, however, if the device was not rooted, the option for a Physical extraction only existed if the phone was rooted.

Since each type of extraction can be only conducted one at a time, the order of extractions began with Physical, File System and then Advanced Logical from the UFED touch which interfaced with *Physical Analyzer* (PA) to conduct the operations of file destination and perform the extractions. The extractions are stored in their own unique folders on the host Windows OS based computer and when opened in PA, creates a file labelled, *'EvidenceCollection.ufdx'*. The (.ufdx) extension is a 'multiple dump file' proprietary to *Cellebrite* that can be opened in PA.

A software add-on to PA is called *Cellebrite Cloud Analyzer* that extracts cloud based digital forensic data from compatible applications from the test device. At the time of writing this paper, *Cellebrite Cloud Analyzer* only supports the *Coinbase* application therefore the cloud-based data from *Coinomi Wallet* nor the *Atomic Wallet* were extracted.

In order to acquire data, the login credentials of the application were required and in addition, required two factor authentication with the login process. The transactions conducted on the *Coinbase* application were extracted and stored as a fourth extraction in the *EvidenceCollection.ufdx* file.

TRANSACTION DETAILS

In Table IV, *Bitcoin* (BTC) and *Dogecoin* transactions were conducted from a cold storage Ledger Nano S (Windows Ledger Live 2.26.1) wallet to the installed applications. The value of the crypto decreases because of transaction fees rewarded to miners as both *Bitcoin* and *Dogecoin* are Proof of Work cryptocurrencies.

Table V identifies the IDs associated with the transactions. These were documented to identify any human readable data found once the test device was extracted using the *Cellebrite UFED Touch*.

*C. IDENTIFICATION AND ANALYSIS*

1) COINOMI TRANSACTION HASH ID ARTEFACTS

Keywords based on the preparation of the extraction were documented and used to search various data sources found in the extraction. The hash IDs and wallet addresses were used as search targets in the *EvidenceCollection.ufdx* project file. The *Coinomi wallet* application was the only application that stored the hash ID as human readable text in the Android file system when it only received cryptocurrency and was found in the Advanced Logical, File System and Physical extraction.

**a. Bitcoin transaction from *Coinbase* to *Coinomi***
Received transactions made by the *Coinomi* wallet application provided the hash ID for the *Bitcoin* and *Dogecoin* wallets and were found in the path of the cache folder (Table IV). The wallet application creates a folder titled by a unique hexadecimal string in the path: /data/com.coinomi.wallet/cache/**f78fc8de58b92a6f**/bitcoin.main/<**HASH ID**>.

TABLE IV
BITCOIN AND DOGECOIN TRANSACTION VALUES

| Transaction | Value |
|---|---|
| BTC: Ledger to Coinbase | .002548 |
| BTC: Coinbase to Coinomi | .002548 |
| BTC: Coinomi to Atomic | .00254 |
| BTC: Atomic to Ledger | .00253 |
| DOGE: Ledger to Atomic | 212.577 |
| DOGE: Atomic to Coinomi | 212.577 |
| DOGE: Coinomi to Coinbase | 210.577 |
| DOGE: Coinbase to Ledger | 210.577 |

***Open Source Blockchain Explorer.*** To simulate an investigative follow-up of acquiring a hash ID from a transaction, open-source tools such as online blockchain explorers provide a wider perspective into a transaction. A characteristic of cryptocurrency transactions is that they are viewable to the public. Online tools such as *blockchain.com* provide information about a cryptocurrency transaction using a transaction, address, or block without attributing it to an actual person (Fig. 4).

TABLE V
BITCOIN AND DOGECOIN TRANSACTIONS WITH ID[7]

| Transaction | ID | Time-stamp (UTC) |
|---|---|---|
| BTC: Ledger to Coinbase | **Hash**:1bfa*******e012a<br>**From**: bc1q*******ejx0<br>bc1q*******n2kr<br>**To**:<br>3DQb*******pTD2<br>bc1q*******qx28 | 14 Jun 2021<br>01:57 |
| BTC: Coinbase to Coinomi | **Hash**:4af2*******8643<br>**From**: 32ti*******K8SW<br>**To**: bc1q*******rnvc<br>(1 of 33 outputs) | 14 Jun 2021<br>03:14 |
| BTC: Coinomi to Atomic | **Hash**:2eeb*******fe73<br>**From**: bc1q*******rnvc<br>**To**: 1EQ6*******Ee4c | 14 Jun 2021<br>03:29:25 |
| BTC: Atomic to Ledger | **Hash**:d232*******f48e<br>**From**: 1EQ6*******Ee4c<br>**To**: bc1q*******3005 | 14 Jun 2021<br>21:49:44 |
| DOGE: Ledger to Atomic | **Hash**:738a*******1f47<br>**From**: DPbi*******r2zw<br>**To**: DMfq*******Awck | 14 Jun 2021<br>03:39:03 |
| DOGE: Atomic to Coinomi | **Hash**:e531*******801a<br>**From**: DMfq*******Awck<br>**To**: DH3T*******BUVG | 14 Jun 2021<br>03:46 |
| DOGE: Coinomi to Coinbase | **Hash**:9aa8*******5269<br>**From**: DH3T*******BUVG<br>**To**: DLou*******cSJT | 14 June 2021<br>03:53 |
| DOGE: Coinbase to Ledger | **Hash**:bf48*******b80a<br>**From**: DDUo*******ze9N<br>**To**: D7JZ*******ERhz | 14 June 2021<br>21:40:57 |

Also from Fig. 4, in the case of the *Coinomi* bitcoin transaction, when the Hash **4af2*******8643** is entered in the search box at *blockchain.com*, it provided information that is consistent with the details of the transaction at the preparation stage, of which the input address was: *bc1q*******rnvc* (Table V).

The record at blockchain.com reveals that the output address of this transaction was: *32ti*******K8SW* but is identified as 1 of the 33 inputs addresses for this hash ID (Fig. 4). *32tidx...* is not unique to the *Coinomi* wallet on the test device since it has transacted 1361 times and has sent/received over 100 million dollars on the blockchain as of July 2021 when referring to the record on *blockchain.com*.

***Cryptocurrency Tracing Analysis.*** *Ciphertrace* is a forensic solutions provider that specializes in Blockchain forensics and provides analytics and tracing of Bitcoin UTXO (unspent transaction output) to find attribution. As seen in the blockchain.com record of the transaction, a wallet address may not be unique to the criminal and the Investigator would have a challenging investigating each of the 33 receiving addresses. Ciphertrace UTXO Whitepaper [23] classified 2 types of cryptocurrency tracing as, "wallet-to-wallet" and UTXO tracing. Using a graphical interface, the tool can ingest hash IDs and addresses. UTXO's are outputs that are returned back from the receiver, back to the sender as 'change', therefore the tracing of the UTXO would be beneficial for the purpose of finding attribution and following the 'money' (Fig. 5). When tracing the hash ID in *Ciphertrace*:**4af2*******8643**, *Ciphertrace UTXO* was tracing back to the sender identifies the source as *Coinbase*. At this point, an Investigator could initiate the judicial process of writing a warrant to obtain records of KYCs and relevant network data held at *Coinbase* since it was also learned that *Coinbase* requires the user to input personal information such as name, email and date of birth.

---

[7] IDs have been partially obfuscated for privacy

TABLE VI
BITCOIN TRANSACTION FROM COINBASE TO COINOMI

| Hash | Path in File System |
|------|---------------------|
| **4af2*******86 43** | **Advanced Logical** Samsung GSM_SM-G965U Galaxy S9+.zip/_data/data/com.coinomi.wallet/cache/f78fc8de58b92a6f/bitcoin.main/**4af2*******8643**<br><br>**File System** Samsung GSM_SM-G965U Galaxy S9+.zip/_data/data/com.coinomi.wallet/cache/f78fc8de58b92a6f/bitcoin.main/**4af2*******8643**<br><br>**Physical** userdata (ExtX)/Root/data/com.coinomi.wallet/cache/f78fc8de58b92a6f/bitcoin.main/**4af2*******8643** |
| From | 32ti*******K8SW |
| To | *bc1q*******rnvc* |
| Timestamp | 13 Jun 2021 23:15 |

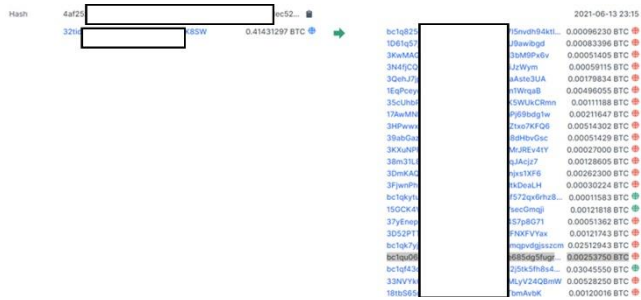

**FIGURE 4.** Transaction information from *blockchain.com* for Hash: 4af2*******8643.

### b. Dogecoin transaction from Atomic to Coinomi

Like with *Bitcoin*, the only artefact found in the controlled group of *Dogecoin* transactions was also the hash ID when the *Coinomi* wallet received *Dogecoin* from the *Atomic* wallet. Like the Bitcoin transaction, the hash ID was found in text path of a cache folder (Table VII): data/com.coinomi.wallet/cache/**f78fc8de58b92a6f**/dogecoin.main/<HASH ID>

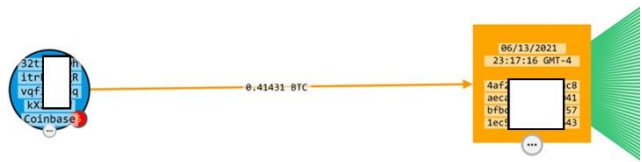

**FIGURE 5.** Ciphertrace graphical interface of tracing from Hash: 4af2*******8643.

TABLE VII
DOGECOIN TRANSACTION FROM ATOMIC TO COINOMI

| Hash | Path in File System |
| --- | --- |
| e531*******801a | **Advanced Logical**<br>Samsung GSM_SM-G965U Galaxy<br>S9+.zip/_data/data/com.coinomi.wallet/cache/f78fc<br>8de58b92a6f/dogecoin.main/**e531*******801a**<br><br>**File System**<br>Samsung GSM_SM-G965U Galaxy<br>S9+.zip/_data/data/com.coinomi.wallet/cache/f78fc<br>8de58b92a6f/dogecoin.main/**e531*******801a**<br><br>**Physical**<br>userdata<br>(ExtX)/Root/data/com.coinomi.wallet/cache/f78fc8d<br>e58b92a6f/dogecoin.main/**e531*******801a** |
| From | DMfq*******Awck |
| To | *DH3T*******BUVG* |
| Timestamp | 14 Jun 2021 03:46 |

***Open Source Dogecoin explorer***. *Dogeblocks.com* is one of numerous *Dogecoin* explorers that provide details about a transaction much like the Blockchain explorers. A search of the Hash ID reveals relevant data such as the date/time of the transaction and both addresses (Fig. 6).

Unlike the Bitcoin transaction, multiple outputs were not associated to the transaction and the Hash ID in the explorer corroborated the recorded data found in the preparation stage (Table V). This type of information would be valuable to Investigators and would provide robust evidence when corroborated with traditional investigative techniques. *Ciphertrace* does not support *Dogecoin* transactions at the time of this paper, therefore Investigators would not be able to followUTXO's in order to trace the transaction to a certain cryptocurrency exchange.

## 2) EMAIL ARTEFACTS

Since no emails were required to set up login access to *Coinomi* or *Atomic* wallets, there were no emails received to the creator of the account that would provide any cryptocurrency transaction details. However, *Coinbase* would require an email address to create an account considering the application is not a proprietary wallet application but also provides exchange services. Regarding *Coinbase* emails, an email was sent to the creator of the account from Coinbase at 'no-reply@coinbase.com'. The subject line of the email that showed, "You just received **0.00254817 BTC**". When Bitcoin was received to the Coinbase application, an email notification was received to the account creator that showed, "You just sent **0.0025375 BTC** to **bc1q*******rnvc**" which documents the input address in the subject line.

For the Dogecoin transactions, two emails were also received to the account creator for both input and output Dogecoin transactions. When Dogecoin was received, the subject line in the email showed, "You just received **210.57749255 DOGE**", and in the subject line for the email for the sent Dogecoin showed, "You just sent **209.13749255 DOGE to D7JZ*******ERhz**".

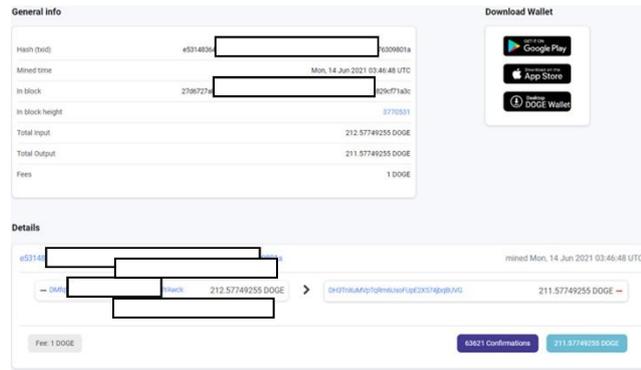

**FIGURE 6.** Transaction information from *Dogeblocks.com* for Hash: e531*******801a.

### 3) WEB COOKIE ARTEFACTS

Samar, V. [24] lays out six parameters for HTTP cookies such as its, "name, value, expiration date, URL path for which the cookie is valid, the domain for which the cookie is valid and whether the cookie should be send using SSL." These parameters were captured with PA inside the Android wallet application file system and will be considered for relevancy in attribution.

**a. Coinomi Cookies**

No cookies were found in the extraction for the *Coinomi* wallet application. The likely reason that no web cookies were found for the *Coinomi* wallet was that while installing the application on the test device, there were no requirements to provide any personal information. Much of the storage of relevant data in order to conduct transactions were found in the cache folders of its file system.

**b. Atomic Cookies**

The Atomic wallet application revealed 16 cookies captured with some cookies of relevance (Table VIII). The web cookie file path for the Atomic wallet application was found in:
**userdata(ExtX)/Root/data/io.atomicwallet/app_webview/Default/Cookies**

The Cloudflare cookie, "_cflb " was created after the installation of the application of the Atomic wallet, but before the times of the Bitcoin (14 Jun 2021 03:29:25; 14 Jun 2021 21:49:44) and Dogecoin (14 Jun 2021 03:39:03; 14 Jun 2021 03:46) transactions conducted on the wallet based on the time stamps.

Three other cookies identified in the Identification stage reveal a value that contains a private IP address range that may not be relevant to Investigators. Although the domains of the cookies displayed in Table VIII are not objective to this paper, the value of the private IP addresses and the domain in the stored cookies provide some insight into the behavior of the wallet application. All three cookies do not have an expiry date and may be considered a persistent cookie. Since it mentions in the Atomic.io terms and conditions that Cloudflare filters all the activity of the wallet application through them, it may be possible for Investigators to lawfully acquire other relevant network records, logs or data from either Atomic or Cloudflare with the creation time stamp, the private IP addresses and the corresponding port numbers for attribution.

| Name | Creation Time | Value | Domain | Expire |
|------|---------------|-------|--------|--------|
| __cflb | 14/06/2021 1:34:34 am (UTC+0) | 0H28vZtZy 6JGHomsa7 GU8NiaZq L8eWyBxR fdXkFSDm v | Bitcoin.ato micwallet.io | 15/06/2021 12:34:44 am (UTC+0) |
| _ef2b1 | 14/06/2021 1:34:43 am (UTC+0) | http://10.0.6 1.221:3001 | nano.atomic wallet.io | None |
| _98548 | 14/06/2021 1:34:42 am (UTC+0) | http://10.2 2.70:21939 | zeus.atomic wallet.io | None |
| _8396a | 14/06/2021 1:34:42 am (UTC+0) | http://10.0. 87.90:8900 | solana.atom icwallet.io | None |

**b. Coinbase Cookies**

There were six cookies identified in PA that also revealed that Coinbase also uses the services of Cloudflare (Table IX). Unlike Atomic, the cookie named, "__cf_bm" has the function of being a 'bot manager'. In summary, there were no human readable values that would immediately be available for attribution, however, understanding how the wallet applications handle their data and web services may provide elements leading to attribution with judicial authorization.

### 4) OAUTH ARTEFACTS

208 password files were acquired in the extraction of the test device with a physical extraction, and in this case, the test device needed root access to acquire this data. A review of the 208 password files indicated that 185 are based on the OAuth 2.0 protocol managed by Google and there were no implementations of OAuth tokens by the installed wallet applications.

Cellebrite Cloud Analyzer (CA) forensically captures data from cloud-based applications installed on mobile devices. CA captured four Bitcoin transactions but did not capture any Dogecoin transactions nor any of the hash/transaction IDs. Since Coinbase was the only supported application by Cellebrite, transactions from Coinomi and Atomic wallet applications were not acquired (Fig. 7).

**FIGURE 7.** Cloud Analyzer of Bitcoin transactions.

In comparison to the data collected at the preparation stage, CA captured three outputs from the Ledger device to the Coinbase application which had a transaction time and date of 14 Jun 2021 01:57 UTC and the fourth output captured by CA indicates an output at 14 Jun 2021 02:00 UTC. These outputs are consistent with at are consistent with the Ledger outputs in the preparation stage for hash ID: **1bfa*******012a** (cf. Table V).

A Ciphertrace analysis of the hash ID identified corroborates Coinbase as the input address for this transaction. The inputs are consistent with the transaction data from CA and corroborates the record on the blockchain. The blockchain explorer identifies the wallet address that received the bitcoin as Coinbase from the Ledger wallet which could assist Investigators in acquiring records of databased evidence through judicial authorization (Fig. 8).

TABLE IX
COINBASE COOKIES

| Name | Creation Time | Value | Domain | Expire |
|---|---|---|---|---|
| __cf_bm | 14/06/2021 5:15:52 am (UTC+0) | 103521a56bbb3c13e10322266bf63edd4f95b20c-1623647752-1800-ASujzNfb8I4XhshgLMp4JQHORm/Y/NrMHOvxx/LoBwb217kuP1ENI/iWFs1fUtmiEzG/X9/vVIMzJGb1kBZUiIM= | .coinbase.com | 14/06/2021 5:45:52 am (UTC+0) |
| __cf_bm | 14/06/2021 1:36:28 am (UTC+0) | 7ba94c661aff0df4e92202a5914b4999db915394-1623634588-1800-AdclxXieaDLhB/uGqBL5RSV+pbM+Nf6tjxBbJkQBbT2FpafvpoBHmpPHq6N2ToXUxDPXjBhJLkJTc+V5G/q5Xygg= | .coinbase.com | 14/06/2021 2:06:28 am (UTC+0) |

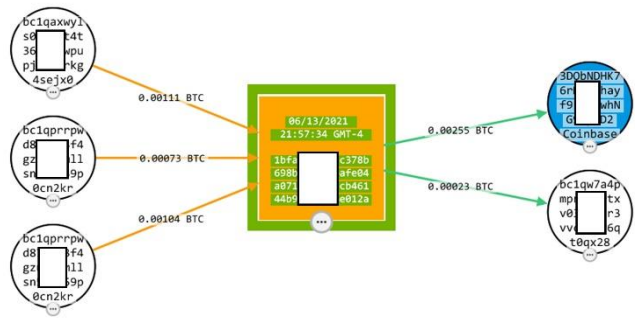

**FIGURE 8.** Ciphertrace graphical interface of tracing from Hash: 1bfa*******012a

## V. EVALUATION AND DISCUSSION

### A. Wallet Application Artefacts

1) CRYPTOCURRENCY REVIEW AND WALLET ARTEFACTS

Previous research for cryptocurrency wallets installed on Android devices have used common examination tools such as Cellebrite UFED and Physical Analyzer (PA) as seen in [20]. Versions and types of applications and operating systems have changed since 2014 which have led to different results from past research into Android wallet applications. The updated Cellebrite proved to be a competent tool that provides pertinent data across the Android file system.

Common digital forensic artefacts were identified in plain sight within the Android file system, but in particular, the Coinomi wallet application stored the hash IDs for the received transactions in the cache folder for the application. When cryptocurrency is received by the Coinomi wallet application, a specific folder using a unique hexadecimal string is created in the file system. In our experiment, a *bitcoin.main* and *dogecoin.main* folder was created in this folder that contained 0 kb files labelled with the hash IDs. This area in the file system would be a valuable area to look for the hash IDs.

Although the function of each wallet application facilitates cryptocurrency transactions, the features of the wallets are unique to one another.

TABLE X
COMPARISON

|  | Coinomi v.8.26.3 | Coinbase v.1.22.3 | Atomic v. 0.75.1 |
|---|---|---|---|
| Email required | No | Yes | No |
| Phone number required | No | Yes | No |
| Purchase crypto with fiat | Yes, through Simplex | Yes | No |
| Artefacts in File System | Yes | No | No |

Each wallet carries its own unique characteristics, but a particular difference between all three is that the Coinbase application is not a dedicated wallet application. An Investigator would need to delineate between a dedicated wallet application and a cryptocurrency exchange application. Cryptocurrency applications acquire KYCs from their users, by requesting their email address, phone number, name and at times copies of personal documents that provided valuable data in the form of emails with the Coinbase application (Table X).

Dogecoin has been a popular coin in mainstream media and not a lot is known about this cryptocurrency in academic research due to the sheer volume of coin types. In support of providing modern research in cryptocurrency forensics, the history and characteristics of Dogecoin were researched in this paper. Dogecoin artefacts in the form of hash IDs were identified in the Coinomi wallet, but attribution fell short of Bitcoin's analysis since Dogecoin technology is not supported by tracing tools such as Ciphertrace. Dogecoin's ledger is public, and Investigators would be able to see transacting addresses and date/times of transactions which may support attribution efforts.

According to Dogecoin's white paper, they have future plans to change the technology of their cryptocurrency from the down up as the demand for sustainable blockchain technology rises. The impact this has on forensic analysis of cryptocurrency wallets would not be large since the actual blockchain technology doesn't directly influence the function of a wallet application. A consideration needs to be made with how criminals utilize cryptocurrency. Based on the wallet applications that were analyzed in this paper, they function similarly by facilitating the transactions of cryptocurrency. There are new emerging technologies such as staking and POS (proof-of-stake) that are challenging older technology such as mining and POW (proof-of-work).

HTTP cookies and OAuth tokens were examined in this paper but provided no relevant data towards attribution. Conducting network analysis in the behavior of cookies and OAuth tokens in modern wallet applications could provide valuable network data such as IPs, addresses, and credentials for attribution.

As observed in our research, each cryptocurrency wallet functioned distinctively in comparison with one another. Coinbase and Atomic wallet application files system did not capture the transaction IDs in their respective cache folders.

2) TRACING ANALYSIS

It would be difficult for Investigators to find attribution using open-source resources such as blockchain.com since an address used by a wallet on a device may be shared by others on the same network. There are a few private vendors that offer cryptocurrency tracing tools such as Ciphertrace that assist in finding attribution.

Hash IDs for Bitcoin and Dogecoin transactions were captured in Coinomi's Android cache folder. As a result, a blockchain explorer search revealed that the output address and timestamp of the transaction were consistent with the transaction details documented during the preparation of the experiment. Although these artefacts do not provide a direct attribution, they would be able to corroborate other evidence in an investigation. For example, during an Organized Crime investigation, transactions of illicit goods recorded from physical surveillance or covert intelligence operations can be corroborated by transaction IDs and time stamps that were acquired from a seized criminal Android device.

The use of Ciphertrace provided a visual map of the hash IDs acquired from the Coinomi wallet application similar to an open-sourced blockchain explorer but was able to attribute the transaction to the Coinbase exchange which was consistent the recorded data at the preparation stage. There are no details provided as to how Ciphertrace acquired attribution to Coinbase which could be challenge for Investigators in sourcing the transaction to a particular exchange. According to [23], their tool analyses UTXO's in a transaction that provides attribution which could be a valuable tool for an Investigator as a supplement to digital forensic tools in order to acquire attribution, but some legal considerations will need to be made for the growth of tools such as these.

### B. Authentication

Human readable data may not be synonymous with authentication, but how cookies and tokens interact between wallet applications and Android is a topic that has not been researched in the context of forensic artefact discovery and attribution. Single Sign On (SSO) has been a topic of research since research was conducted by [24], and with the advancements in authentication, it would be valuable to consider while investigating cryptocurrency since login credentials that include usernames and emails would assist in attribution.

All but Coinomi revealed HTTP cookies in the examination and there were no OAuth/password tokens attributed to any of the installed wallet applications. All the tokens were managed by Google within Android. Tools such as Cellebrite Cloud Analyser provides access to cloud-based accounts such as Coinbase, but the Investigator was required to be in possession of both the user credentials and the two-factor authentication destination. The Coinbase and Atomic cookies that were examined provided insight that the applications use a third party to serve the cookies which didn't provide any human readable artefacts. The cookies were identified as session-based cookies that provided creation and expiry dates that may assist in attribution but would not be a robust avenue for discovery. OAuth and cookie forensics have been subject to various research articles and dissertations but have not been popular with the topic of cryptocurrency.

Regarding Cloudflare cookies for Coinbase, Cloudflare's website explains the details about this cookie that has a purpose of, protecting the client's site from bad bots[8]. Acquiring network data records from Coinbase, Cloudfare may be able to assist Investigators with attribution.

Regarding Cloudflare cookies for Atomic, Cloudflare is an internet solutions provider that offers the service of data and http requests worldwide for its clients. The "_cflb" cookie is a 'Cloudfare Load Balancer" that is characterized as a session cookie which has an expiry and, in this case, expired 35 hours after it was created and has a function to track the "associated origin web server"[9]. Load balancing distributes the workload of traffic to a webserver by replicated servers in the backend in the event of server failure. Cloudfare indicates that they include the use of these cookies for the client to optimize the user's experience. In their terms and conditions and in this case, Atomic.io posted the usages of cloudflare cookies in their policy[10].

If the services of Cloudflare, "filters all the traffic through" the Atomic application, this may be an avenue of evidence for Investigators to consider. Content cannot be removed off a website or application by Cloudflare, but it is unknown if there are logs or other pertinent artefacts of evidence available for Investigators with judicial authority held by them.

---

## C. Cryptocurrency Tracing Tools

The use of a cryptocurrency tracing tool is relatively new to digital investigations and Ciphertrace was used to determine attribution once an examination of forensic artefacts was conducted on the test device.

Cryptocurrency tracing may require modernized legal considerations since the source of the tracing technology exists in the backend of the tool's operational structure that may not be readily available to Investigators as a reliable source. As an example, the tool provides attribution to an exchange, but does not provide how it came to that conclusion. In this experiment, the exchange provided by Ciphertrace was consistent with the data recorded in the preparation stage. Although it is valuable information that may lead Investigators to seek judicial authorization to acquire records from Coinbase, it may be difficult to source and articulate the accuracy in a judicial authorization.

## D. Comparison with Other Cryptocurrency Wallet Research (Table XI)

TABLE XI
RESEARCH COMPARISON OF RECOVERED ARTEFACTS

| | Montanez [20] | Haigh [25] | Li [1] | Van Der Horst[18] | Thomas [26] |
|---|---|---|---|---|---|
| **Year** | 2014 | 2019 | 2020 | 2017 | 2020 |
| **Real World Transactions** | Yes | Yes | No | Yes | Yes |
| **Hash/ Transaction ID** | Yes (ADB pull) | Yes (XML and SQLite Databases) | Yes (Tampering Application) | Yes (File System and Memory Image) | Yes (Process Memory) |
| **UTXO** | No | No | No | No | No |
| **Cookies** | No | No | No | No | No |
| **Email Artefacts** | No | Yes (in application code) | No | No | No |
| **OAuth and/or Passwords** | No | Yes (plaintext password) | Yes (Accessibility Service) | No | Yes (Passphrase in Process Memory) |

The approach of [20] to cryptocurrency wallet analysis acquired an extraction using a virtual emulator Genymotion using the Android Debugging Bridge (ADB) pull command that resulted in data related to the active session when the pull was made and erased after the emulator was shut down. While this is a unique method of forensic acquisition, it provided no evidence that would lead towards attribution. The tools from Cellebrite forensic tools provided a more robust analysis of the test device. The disadvantage of this is that third party vendors such as Cellebrite are costly and may not be readily available for use to some law enforcement agencies. Since Montanez, A (2014) did not conduct any cryptocurrency transactions in the research, very little data was acquired when focusing on just the installation of cryptocurrency wallet applications on a mobile device. To acquire a robust forensic acquisition in a testing environment, conducting a real-life scenario such as conducting cryptocurrency transactions with the subject wallets would influence the behavior of the applications that would result in acquiring artefacts for attribution.

Haigh et al. [25], conducted an analysis of 7 different cryptocurrency wallet applications and had a succinct approach to acquiring forensic artefacts from Android cryptocurrency wallets. Haigh et al conducted cryptocurrency transactions on an Android device which resulted in the discovery of plaintext passwords and transaction details viewable in plain text from identified databases. Although an objective for a forensic acquisition of a cryptocurrency wallet can be to identify private keys and seed phrases which may not likely be stored locally on a device, Haigh et al concluded that six of the seven applications acquired transaction history which would assist in attribution. The advantage of [25]'s approach introduces a "Trojan proof-of-concept" that is intriguing since it places the investigator in an adversarial role which resulted in positive results for forensic artefact acquisition. When reviewing the results of their 7 different wallets, it was apparent that each wallet behaved differently from one another. For example, it was discovered that Coinbase did not store any private keys on the device, Bitcoin Wallet contained all wallet information such as private keys. This supports the need for further analysis and reviews of other cryptocurrency wallets that would assist with the goals of attribution in investigating cryptocurrency transactions. There are contrasting perspectives with the research approach with our work in this paper, yet the results of forensic artefact discovery may be similar. Using common forensic tools, we discovered cryptocurrency artefacts in plain sight with less superior techniques.

Authors in [1] were studying the risks of Android-based cryptocurrency wallets by using an adversary model to analyze attacks originating from the Android operating system. They provided a view from a defensive position while Haigh et al. [25] provides their "Trojan proof-of-concept" from an offensive position. These perspectives can be a valuable consideration for future research to provide a wider perspective into digital operations by criminals.

Authors in [18] analyzed Bitcoin clients within the Windows operating system was focused on process memory investigation. Although Windows Bitcoin clients differ that Android wallet applications, the approach used to investigate the process memory in an Android wallet application would be beneficial. They acquired memory images from a virtual machine running in various states. This resulted in the acquisition of Bitcoin data such as public keys and addresses where binary values were used to search the memory for cryptocurrency wallet artefacts. Memory encryption could hinder analysis and would take more time for an investigator to acquire attribution of a cryptocurrency transaction which supports the analysis of human readable and plain sight artefacts.

Thomas, T et al. [26] presented forensic research on Ledger Nano X and Trezor One cryptocurrency hardware wallets using a tool called FORESHADOW, also known as Memory FOREnSics of HArDware cryptocurrency Wallets.

Unlike software mobile wallets presented in this research, hardware wallets are physical devices used as a 'dedicated private key storage devices' that uses cryptography to isolate private keys from 'any part of the device that is externally accessible. While being partially air gapped from the internet, hardware wallets interact with software clients[11].

An analysis of process memory was conducted in the interaction and presented various forensic artefacts when a hardware wallet was used to conduct cryptocurrency transactions.

An analysis The LedgerLive provided transaction history, public keys, metadata and public keys, while the Trezor Wallet client provided transaction history, wallet addresses, metadata, public keys and an extractable passphrase.

## VI. CONCLUSION

Cryptocurrency artefacts can be found in plain sight on an Android file system. An application such as the Android Coinomi wallet provided hash IDs of received transactions in its cache folder. This would prove to be a relevant source of attribution for an investigation involving cryptocurrency. Email artefacts related to the Coinbase wallet application provided wallet addresses of transactions. Since the application cache provides the user with convenience in accessing the application, the hypothesis that forensic artefacts exist in plain view since mobile operating systems aim to create a better user experience proved to be true

Although each Android wallet application shares the same function of providing users the ability to trade and purchase cryptocurrency on a mobile device, each application has its own unique features. As displayed in table 8, the features of each wallet are unique which translated into different results for the purpose of acquiring forensic artefacts. The hypothesis that each wallet application would provide unique features and different results were accurate.

HTTP cookies and OAuth tokens were examined in this paper but provided basic information such as date and time stamps that may corroborate a far-reaching view of when cryptocurrency transactions were made according to the creation and expiry date and time. Although there were not significant avenues of attribution, the technology each wallet uses to manage their HTTP cookies provided some insight into third party vendors such as Cloudflare which claim that all the data (not content) flows through their computing network. This would be a relevant research topic on how Cloudflare technology can provide attribution in a cryptocurrency investigation. No OAuth tokens or passwords were managed by the wallet applications but by Google.

Cryptocurrency tracing tools are effective and corroborated the data recorded for the transactions and the preparation stage. The hash ID found in the Coinomi wallet application file system provided no attribution using blockchain explorers, but Ciphertrace accurately identified Coinbase as the output exchange in the transaction.

---

[11] Ledger Live and Trezor Wallet software clients

Are cryptocurrency tracing tools effective and would it be a competent tool in support of digital forensic techniques for the purpose of attribution? A hypothesis to consider is although blockchain explorers transparently reveal transaction using open-source tools, the features of tracing tools providing transaction intelligence will need to be corroborated and need lawful consideration when used as a source for evidence gathering.

In conclusion, cryptocurrency investigations are not a new phenomenon for traditional law enforcement investigations, but the rapid growth of it requires an Investigator to rely on modern research. Common definitions associated with cryptocurrency such as Bitcoin and Blockchain have now become essential to the traditional law enforcement Investigator. Blockchain technology has multiplied to a variety of use cases such as staking and asset management that make for a small percentage of growth in this ecosystem and was not considered in this paper for research. In addition to sophisticated cryptocurrency forensic techniques, human readable data exists on Android devices.

### *Limitation of this Research*

The evaluation of this research only focused on the artefacts of Android Samsung phone however the proposed approach could be applied to other Android devices. In addition, this approach could also be extended to deal with non-Android devices such as iOS-based devices. Besides, currently the approach only analyzed the artefacts related to Bitcoin and Dogecoins, but it could be adopted for analyzing other cryptocurrencies, it is subject to the future work.

### *Possible Future Research*

Bitcoin and Dogecoins are a few of thousands of cryptocurrencies available on the market. While popular, there is a move towards a more energy efficient and sustainable cryptocurrency. The proof-of-work (POW) consensus mechanism regulates the processes of transactions with Bitcoin and Dogecoin, but the proof-of-stake (POS) consensus mechanism is also required. There has not been any research focusing on the acquisition of forensic artefacts associated with POS cryptocurrency and related mobile cryptocurrency applications and it is subject to our future work.

Cloudflare cookies were identified in both Atomic and Coinbase wallet applications in this paper. There were no human readable artefacts identified in the examination of the test device. There is little to no research on this cookie offered by a third party, which operates as a computing service for the wallet applications. Future research in this area may provide valuable forensic and security insight into cryptocurrency wallet applications and other mobile based applications.

Besides, on March 11[th], 2022, coinmarketcap[12] added almost 30 new cryptocurrencies within 24 hours. This demonstrates the ever-growing array of cryptocurrency options available to criminals. Deliberate attention to researching new variations of cryptocurrency artefacts is essential to progressive criminal investigations.

---

12 https://coinmarketcap.com/new/